\documentclass[11pt,compsoc]{IEEEtran}
\usepackage[utf8]{inputenc}
\usepackage{times}
\usepackage{url}
\usepackage[colorinlistoftodos]{todonotes}

\title{The Difficulty in Scaling Blockchains: A Simple Explanation}
\author{
    \IEEEauthorblockN{Maarten van Steen} \\
    \IEEEauthorblockA{University of Twente, The Netherlands \\ 
    m.r.vansteen@utwente.nl} \\
    \vspace*{6pt}
    \IEEEauthorblockN{Andrew A. Chien} \\
    \IEEEauthorblockA{University of Chicago, USA \\
    achien@cs.uchicago.edu} \\
    \vspace*{6pt}
    \IEEEauthorblockN{Patrick Eugster} \\
    \IEEEauthorblockA{Universit\`{a} della Svizzera italiana (USI), Switzerland \\
    patrick.thomas.eugster@usi.ch}
}
\date{}

\begin{document}

\maketitle

\begin{abstract}
    Blockchains have become immensely popular and are high on the list of national and international research
    and innovation agenda's. This is partly caused by the numerous interesting applications, combined with the
    promise of full decentralization and high scalability (among others). Keeping that promise, however, is
    technically extremely demanding and has already required substantial scientific efforts, which so far have
    not been overwhelmingly successful. In this paper, we provide a laymen's description of what may turn out
    to be a fundamental hurdle in attaining blockchains that combine scalability, high transaction processing
    capabilities, and are indeed fully decentralized.
\end{abstract}


\section{Introduction}

Since the introduction of Bitcoins we have seen a steady increase of interest in blockchain technology. Its popularity is often attributed to its decentralized nature (thus avoiding the need to trust a single or a few central entities), its openness (anyone can, in principle, join a blockchain system), its transparency (transactions are public, and so is the process), as well as its immutability (once committed and validated, a transaction cannot be altered)~\cite{zheng.z2017,wu.m2019}. At the same time, there are numerous challenges. For one, much effort is put in making blockchains robust against a myriad of security attacks. Also, the openness and transparency bring issues related to privacy. Meanwhile, a myriad of applications other than cryptocurrencies have been proposed. Most, if not all of these applications exploit the fact that one can avoid relying on a trusted entity, be it a governmental body, a bank, or an educational institute, to name few.

One of the problems with blockchains is that the underlying technology is not that simple to explain, let alone easy to understand. In fact, we argue that a significant level of technical expertise is needed to grasp the many underpinnings of blockchains and that it requires some serious studying even by experts to understand all the details. As a result of these intricacies, there is now a huge gap between understanding applications of blockchains (which is relatively simple) and understanding what is going on under the hood (which is relatively difficult). That gap becomes troublesome when adopters of blockchain applications are confronted with unanticipated limitations that stem from the blockchain technology itself, but which were simply not well communicated. The lack of proper communication is often not intended, but comes from the fact that much of the technology is not sufficiently or completely understood to allow for an explanation to a non-expert.

One such source of misunderstandings, and the focus of this paper, is that of scalability, which is generally expressed in terms of transaction processing capabilities. A recent survey by Deloitte shows that there is a firm belief that eventually blockchains will indeed scale~\cite{deloitte2019}. However, when digging deeper into the issue, matters are not that simple and there is reason to believe that building scalable blockchains may require sacrificing one or more of its other attractive features. To clarify, blockchains systems are often presumed to offer the following four properties (see also~\cite{bashir.i2017}):

\begin{description}
    \item[\textbf{[GC]}]Global consensus on validated transactions.
    \item[\textbf{[TP]}]Validation of transactions without the need for a trusted third party.
    \item[\textbf{[SC]}] Scalability in terms of transaction processing capacity.
    \item[\textbf{[SP]}] Scalability in terms of number of participants.
\end{description}

As mentioned, discussions on scalability are often limited to transaction processing capacity, whereas the combination of open membership, lack of a trusted entity, and lack of any centralized component is associated with what we refer here to as scalability in participants.

Trivially, if we drop property \textbf{GC}, we can easily realize the other three: simply let any participant locally validate transactions. Of course, such validations will generally be meaningless. When sticking to \textbf{GC} and dropping \textbf{TP}, there are known solutions that achieve high throughput, and as we shall see, progress is being made when it comes to \textbf{SP}, although progress is slow and uncertain. Sticking to \textbf{GC} and \textbf{TP} allows us to establish \textbf{SP}, yet there are inherent limitations in current proposals that will prevent us from establishing \textbf{SC}.

We postulate that there is currently no solution for simultaneously establishing \textbf{GC}, \textbf{TP}, \textbf{SC}, and \textbf{SP}. In fact, there are good reasons to doubt that a combination of these four properties will soon be possible at all. 

In this paper, we concentrate on explaining the scalability problems of blockchains. We target readers lacking (deep) knowledge of blockchains, yet who are looking at various applications of blockchains for their own field of expertise. Also for this reason, we  simplify matters as much as possible, to the extent that simplifications do not lead to incorrect reasoning. We note that various and often excellent surveys exist of blockchains, some of which are quite comprehensive, such as~\cite{dinh.t2018,wu.m2019}. Overview articles concentrating on cryptocurrencies include~\cite{tschorsch.f2016,holub.m2018}. Scalability is explicitly addressed in~\cite{croman.k2016}, with a strong focus on transaction processing capacity. We stress that although we are quite pessimistic about the scalability of blockchains, we also mention few works in this paper that may eventually prove that more optimism is in place.

\section{Blockchain 101}

The key issue in many transaction systems is that a transaction is validated, effectuated, and subsequently stored for various auditing purposes. For example, Alice may decide to create a transaction stating that she transfers \$10 to Bob's account. Normally, she would go to a bank where she would have to sign the transaction to prove that she really wants it to be carried out, the bank will check whether she has enough credit, whether Bob is eligible for receiving the money, and assuming everything is fine, will subsequently transfer the money. A record of the transaction is kept for all kinds of auditing purposes.

The bank operates as a trusted third party. In the case of blockchains the idea is that a transaction is digitally signed after which we let ``the network'' (generally a collection of validators) handle the transaction's validation, effectuation, and storage. To this end, validators explicitly pick up fresh unvalidated transactions from the network. A number of these transactions are grouped into a block (of which the size and structure is protocol-dependent), after which they are validated. Validation will generally depend on previously validated transactions, which are all available in the form of a publicly accessible blockchain. If everything goes well, the validator securely protects the group of validated transactions against any modifications, and appends the now immutable block to that chain of other blocks with validated transactions. 

An important observation is that there is logically only a single chain of blocks with validated transactions. Each block is immutable, in the sense that if an adversary decides to modify any transaction from any block in that chain, the modification can never go unnoticed. Securely protecting blocks of transactions against modifications, but also securely appending a block to an existing list are well-understood techniques. The immutability of a block makes it an ideal fit for massive replication: it will never be changed anyway so someone may just as well store it locally to make verification as simple as possible. Effectively, the logically single chain of immutable blocks may be physically massively replicated across the Internet.

The problem is that we want to employ a large number of  simultaneously operating validators (the basis of decentralized trust and scale). As a result, we need to figure out who is allowed to append a block of validated transactions to the existing chain. Appending such a block means that there is global consensus on fulfilled transactions. It is therefore important that we also reach consensus on which validator can move ahead. All others will have to do their validation over again, as their transactions may be affected by the newly appended ones and thus may need to be revisited.

Deciding on which validator can move ahead requires distributed consensus. As we shall see, if we want to have many transactions per time unit, we need a trusted third party, but also in that case the number of validators will need to be limited. Many validators can be supported by avoiding a trusted third party, but this will necessarily lead to relatively long waiting times before reaching consensus, thus seriously limiting the processing capabilities of the transaction system as a whole. Further, this will lead to significant energy challenges in scaling.  Indeed, it is where deciding who can append a block to the blockchain where scalability is potentially at stake. 

\section{Distributed consensus}

Let us now concentrate on how validators can reach consensus on who is allowed to append a block of validated transactions to the blockchain. There are essentially two main approaches to accomplish consensus, which we describe next. A more thorough discussion on consensus in blockchains can be found in~\cite{bano.s2017}.

\subsection{Racing-based protocols}

The first method for reaching consensus is essentially letting the validators run a computational race. The winner is allowed to append a block. Computational races are based on solving a computationally difficult problem, i.e., a problem for which a solution is known to exist, but also that it takes a lot of compute power to solve. The type of problem used for blockchains is based on what is known as \textbf{hashing}. A hashing function takes as input a (possibly large) data set and produces a fixed-length string of bits, typically of length 1024, called a \textbf{hash}. A cryptographically well-defined hashing function has a number of important properties~\cite{ferguson.n2003}:

\begin{enumerate}
    \item Computing the hash of a given data set is relatively easy, i.e., it does not require significant computational efforts.\label{prop:computation}
    
    \item However, given a specific hash, it is computationally very difficult to find a corresponding data set with the same associated hash. \label{prop:inversion}

    \item With very high probability, any two different input data sets will lead to two seemingly unrelated different hashes. Even if the two data sets are minimally different, their associated hashes will most likely be very different.\label{prop:collision}

\end{enumerate}

The third property is used to securely protect blocks of validated transactions against adversarial modifications: even the change of a single bit will not go unnoticed. Combined with the second property, it becomes virtually impossible to modify a block in a meaningful way such that the modified block's hash is the same as the original block's hash.

To set up a race between validators, each validator computes the hash of its block of validated transactions. Note again that a hash is technically just a fixed-length string of bits. Let's call the computed hash over a block a \textbf{digest}. The validator is then required to find an \emph{additional} bit string, called a \textbf{nonce}, such that the hash computed over the digest and the nonce when taken together produces a bit string with a predetermined number of leading zeroes. Given property \ref{prop:inversion} of hashing functions, we know that it is computationally very difficult to find a data set that matches a given hash. In our case, we have as input the hash of the block of validated transactions (i.e., the digest), and are now required to find a bit string (i.e., nonce) that, taken together with the digest, has an associated hash that starts with a given number of zeroes. This is computationally very difficult. In essence, each validator simply needs to go through a most likely lengthy trial-and-error process of generating a nonce and checking whether that nonce, combined with the digest, will lead to the required result.

By controlling how many leading zeroes the outcome should have, we essentially control the difficulty of the computations. For example, with just 1 leading zero, there is essentially a 50\% chance that a generated nonce will lead to the desired result. Demanding 2 leading zeroes reduces this probability to 25\%, 3 leading zeroes to 12.5\%, and so forth. With 64 leading zeroes, which is current practice, the chance that an arbitrarily chosen nonce will do the trick is a mere 0.00000000000000005\% (that's 17 zeroes after the decimal dot). Put differently, a validator will on average need to check about 18 billion billion nonces to find one that leads to the desired result. Using dedicated supercomputers, this takes about 10 minutes. It would take an average laptop about 100 years. 

With increasing hardware capacity, the time it takes to find a nonce will drop. However, in racing-based consensus systems, the difficulty of finding a nonce is deliberately controlled in such a way that a race will have an expected duration. For Bitcoin systems, the duration is approximately 10 minutes. If races tend to become shorter, then the difficulty for finding a nonce is increased. If races turn out to be too lengthy, the difficulty is lowered. 

There is a reason for properly setting the expected duration of a race. Suppose we have very lengthy races. This means that validators would have to do a lot of work to find a proper nonce. The good part of the story is that the chance that two validators  find a nonce at more or less the same time is relatively small (although this depends, of course, on how many validators participate in the race). As a consequence, the chance that there will be only one winner who will have enough time to successfully broadcast its victory to all others is relatively high. 

The downside of the story, however, is that with lengthy races the transaction processing capacity may be too low. Suppose that we have a race duration of 10 minutes and an average number of 2500 transactions per block (which are common values for Bitcoin). This means that, effectively, the system can handle about 4 transactions per second. Increasing the number of transactions per block is an obvious path to follow and has been subject to much debate. Assuming that transactions have already been distributed to all the nodes, then a validator need merely distribute the \emph{transaction identifiers} in a block. Ignoring some overhead, with an identifier size of 32 bytes and a block size of 1 MB, we would be able to increase the transaction rate to 52 transactions per second. Only when also moving to a block size of 32 MB would it be possible to come to close to transaction rates of commercial financial systems. We are  still far away from such figures.

On the other hand, assume we would have very short-lasting races. In that case, two validators may more easily concurrently find a proper nonce, both will declare victory, append their block to the blockchain (which was already massively replicated), and before other validators had the chance to stop their work, we would find ourselves with different copies of the same blockchain. There are different ways to correct for these concurrent versions of what is supposed to be a single blockchain, but obviously it comes at the price of invalidating potentially many transactions, and rolling back to a valid state. Of course, the advantage of a smaller race duration is that, in principle, we can handle more transactions per time unit. But when using the current statistics on the Bitcoin network, the transaction rate will remain relatively low. The current network has approximately 10,000 nodes for which it takes some 5 seconds to distribute a block of transaction information to at least 90\% of those nodes~\cite{neudecker.t2018}, possibly after some optimizations.\footnote{See also \url{https://dsn.tm.kit.edu/Bitcoin/}.} We could, theoretically, go for a race duration of 5 seconds in which we could handle 2500 transactions. That is still a mere 500 transactions per second, which is still far less than what commercial systems achieve.

Many of these back-of-the-envelope calculations are still optimistic and ignore additional costs for retransmissions or corrections for prematurely appended blocks in the case of concurrently winning validators. Even if we assume that distribution speeds can go up significantly, we can only conclude that property \textbf{SC} cannot be easily met if we want to maintain \textbf{SP}. Indeed, lowering the number of participants will certainly contribute to speeding up the transaction rate, but this means we will have to let go of property \textbf{SP}.

Racing-based schemes also have significant scalability issues in terms of energy cost, in the face of preserving blockchain openness properties. The Bitcoin community has shifted over time to radically more energy-efficient hardware-implementations of hashing. In fact, over the past 15 years, the energy per hash computation has dropped by nearly 1 million-fold, a total progress comparable to thirty years of Moore's Law progress~\cite{Fuchs.a19}. In the most recent hardware the energy per hash is approximately 160 picojoules. This is truly remarkable and for comparison Nvidia's machine learning optimized DGX-2 consumes approximately 5 nanojoules per FP16 operation! This radical energy efficiency has been achieved by the aggressive engineering and adoption of customized accelerators that exploit custom circuit designs and the massive numbers of low-cost transistors available to deliver this dramatic increase in energy efficiency.\footnote{There is a corresponding increase in performance per device, but that is not our focus here.} Yet, despite that 1 million-fold increase in hardware efficiency, the power consumption of the Bitcoin network continues to grow rapidly and the power per transaction is a staggering 644 KWh.\footnote{See \url{https://digiconomist.net/bitcoin-energy-consumption.}}  At this level, with just 20 Bitcoin transactions you could match the per capita electricity consumption of the United States. We believe this cost is due to scaling flaws inherent to racing-based protocols.

The challenges in balancing race duration, block size, and transaction rate also create fundamental obstacles to the energy efficiency of racing-based protocols whilst maintaining openness and decentralized trust.  Intuitively, as scale and distribution is increased, the cost of coordination and validation will grow (consensus), thus increasing the energy cost per transaction.  If network transaction processing capacity is limited, as discussed above, and the network continues to grow (openness), then the energy per transaction will grow in proportion to the size of the network.  This arises from blockchain costs for distribution and replication which must grow in proportion to the size of the network.  Yet these costs grow further from limits in effective coordination around races: the more validators we have, the more complex is the coordination.  The limits around short blocks are a good example of this -- already effectively limited by a lower bound of perhaps 5 or 10 seconds to reach consensus among 90\% of the participants, any further lowering of this bound will force validators to roll back because they wrongfully assumed to have won the race. The effect is an increased waste of computations due to roll-back efforts (energy). It is worth noting that incentivized validation schemes such as blockchains further suffer from  ``idling on:'' validators are typically incented to continue to work (consuming energy) even if there are no useful transactions to perform.  Note that even Ethereum, Bitcoin's ``more efficient'' sibling, still has a staggering 30 KWh energy cost per transaction.\footnote{See  \url{https://digiconomist.net/ethereum-energy-consumption.}}  In contrast, conventional protocols based on centralized trust can execute thousands of transactions per KWh.

The protocols used for blockchains are not able to maintain the four acclaimed properties, and the limits we have analyzed also imply fundamental challenges in energy scaling, so we conclude that racing-based protocols for blockchains need to be carefully rethought as a solution for building blockchains. In addition, we take the standpoint that computational races are not an acceptable instrument for designing distributed systems. 

\subsection{Talking-based protocols}

Racing-based protocols have the advantage of being open: anyone who wants to operate as a validator has the opportunity to do so without further ado. Simply join a race, try to win, and announce victory when a proper nonce has been found. In talking-based protocols, consensus is reached within a closed, known group of validators. As a result, to become a validator, one has to be admitted to the group. 

Reaching consensus in a closed group of validators is a well-known problem that has received considerable attention in the distributed systems community~\cite{cachin.c2011}. It belongs to the foundations for developing dependable distributed systems. The essence of the associated protocols is that all participating nodes exchange enough information with each other so that in the end, there is consensus on which action to take. Formally, consensus means that all nonfaulty nodes will have reached agreement on the next step to take, such as which validator is allowed to append the next block to the blockchain.

If we could assume that messages are never lost and validators never fail, then reaching consensus in a closed group is very simple: each validator shares its information with the others and by some deterministic rule that everyone follows, the validator that can go next is selected by everyone else. 

However, messages do get lost and validators do fail or may even be corrupted. Talking-based consensus protocols are designed to operate when many things can go wrong. To this end, they generally operate in a series of phases. Normally, for each such series a \textbf{leader} has been elected to coordinate the actions and to drive the protocol from one round to another. Leader election is relatively simple for closed groups, and even when leaders fail or when mistakenly more than one leader is operating, talking-based protocols will simply elect a new leader (and start over again), or abandon the work that was coordinated by a redundant leader.

To keep matters simple, a leader first collects the status from the other validators during a \textbf{prepare} phase. The prepare phase is used by the leader to collect enough information from validators to be able to formulate a proposal for which block is to be appended to the blockchain. During the \textbf{pre-commit} phase, this proposal is sent out to all validators. The important aspect here is that the leader collects enough acknowledgements to know for sure that a majority of the validators will agree on the next block to be appended. Once that has happened, it sends out the final decision as part of the \textbf{commit} phase. 

What makes most talking-based protocols intricate is that we need to make sure that all nonfaulty validators eventually have the same information. To this end, validators not only communicate with the leader, but also with each other. By doing so, each validator essentially needs to collect messages from \emph{enough} other validators to conclude whether or not (1)~there may be consensus on what to do, and (2)~that there is a majority that came to the same conclusion. An important drawback of these schemes is that because of the extensive communication between all participating validators, the communication costs in terms of the number of messages that need to be exchanged grows superlinearly with the number of validators. In other words, it is difficult to establish property \textbf{SP}. 

Talking-based protocols have shown to scale well in terms of transaction rates, thus meeting property \textbf{SC}. It is unclear to what extent they can actually support many validators, although recent work on the Libra\footnote{Now called Diem.} protocol that is to be used by Facebook indicates that communication costs may be able to grow only linearly with the number of validators. However, it is clear that we need to trust the group of validators. Effectively, talking-based protocols can be used to set up a \textbf{consensus-as-a-service}, yet that service does need to be trusted, thus violating property \textbf{TP}.

We conclude that, just like racing-based solutions, talking-based approaches cannot live up to the four claims made for blockchains. 

\section{Outlook}

Not accepting computational races for designing any blockchain brings us to the point that we need to reach consensus by a talking-based protocol. An important first step by designers of racing-based solutions has been made by the Ethereum system which plans to replace racing with a system that is based on how much a participant is willing to put at stake for reaching consensus. But even here, it has been suggested to run a computational race from time to time to ensure a certain level playing field.

When it comes to more traditional talking-based protocols, recent work has concentrated on scaling up the number of participants while maintaining high transaction processing capacity. There are essentially two issues to deal with. First, is it possible to design an \emph{open} protocol, i.e.\ a protocol in which a participant need not know about all other participants? Second, is it possible to design a protocol that is scalable in the number of participants?

Designing an open talking-based consensus protocol has been the subject of some research for over a decade. As described in recent work~\cite{alchieri.e2016}, such consensus can be achieved in a more or less hierarchical fashion. The authors distinguish a sink of participants who know each other and essentially form a closed group, from participants who do not have knowledge about the entire group. The latter communicate and propagate information among the participants known to them, and are assumed to be connected to the sink. The latter run a known talking-based consensus protocol as soon as all information is available to them. The result is disseminated to all participants. Although the solution partly solves scalability problems, the protocol assumes that appropriate measures have been taken to safeguard against \emph{sybil} attacks by which the identity of a participant can be compromised~\cite{douceur.j2002}. Also, to what extent the protocol affects the transaction processing capacity is unknown.

A number of recent works on scaling up the number of participants in talking-based consensus protocols should be mentioned. In Federated Byzantine Agreement (FBA), as laid down in the Stellar protocol~\cite{mazieres.d2015}, the essence lies in having a participant note that it will agree provided a number of designated trusted other participants agree as well (a so-called quorum slice). Two participants can reach agreement if their quorum slices have some participant in common: only then can information be disseminated across quorum slices. Consensus is reached when a participant is in a quorum, and so is its designated quorum slice. In effect, FBA achieves higher scalability than traditional consensus protocols by avoiding that \emph{all} participants need to talk to \emph{all} other participants. FBA is also open, but does require that participants trust members in their designated quorum slice. With an acclaimed transaction processing capacity of 1000 transactions per second, FBA would seem to come close to meeting our four requirements. However, practice dictates otherwise. As recently pointed out, not only is the Stellar network, an instance of FBA, relatively small, it is also highly centralized~\cite{kim.m2019}. In fact, the network needs to have a number of participants that can be trusted to behave well, but also needs to be configured in such a way that cascading failures can be mitigated. By and large, it is unclear to what extent FBA can practically combine \textbf{GC}, \textbf{TP}, \textbf{SC}, and \textbf{SP}.

Another line of recent work argues that for the specific purpose of asset transfers between wallets as targeted by cryptocurrencies consensus is not actually needed~\cite{rachid2019}. More specifically the authors show that double spending and inconsistencies can be avoided by enforcing only causal order among transactions, so there is no need for a totally ordered chain of blocks.  This mirrors the approach chosen in the IOTA distributed ledger\footnote{\url{https://www.iota.org}}, which induces only a partial order among blocks. While thus avoiding \textbf{GC} may lead to more efficient solutions specifically for asset transfers, this approach may not be viable for many other blockchain applications which actually do require a total order. 

Very recently the LibraBFT protocol has been launched~\cite{bano.s2019}, which is used as the basis for Facebook's cryptocurrency. LibraBFT is based on HotStuff~\cite{yin.m2019}, one of the first BFT protocols to have  managed to bring down the practical communication complexity in reaching Byzantine agreement. Technically, LibraBFT outperforms many existing consensus protocols used for blockchains, where initial evaluations have shown to scale well to more than 100 participants. What remains is that even under these circumstances, LibraBFT is not open and indeed, one needs to trust those participants to do their job well. In other words, we are still confronted with a violation of \textbf{TP}. The discussions on the role of Facebook with respect to its cryptocurrency illustrates that this issue of trust is not just technical by nature but also includes a significant social-economic component.

The problem of reaching consensus is simplified if we can assume that participants securely draw a random number and agree to let the one with the highest number move forward as the winner. This principle has been proposed in the proof-of-luck system, which employs a trusted execution environment such as SGX~\cite{milutinovic.m2016}. However, regardless whether such an approach can be readily made robust, we are still confronted with the limitations of reaching consensus among a very large group of participants in an open environment, as we sketched above. Also note that in the proof-of-luck system, the authors assume epochs lasting at least 15 seconds instead of our (very) optimistic 5-second durations.

In this sense, much more promising are proposals in which not \emph{all} validators, but a subset are endorsed to make a decision. A fundamental design principle is to ensure that all validators have an equal (or purposefully designed weighted) chance of being elected into an endorsement set. This is the essence in Algorand~\cite{gilad.y2017} and xxBFT~\cite{praxxis.2019}, which use a randomly selected set of endorsed validators each time a new block needs to be appended to the blockchain. Using, for example, verifiable random functions in combination with a common, agreed-upon initial seed, it becomes possible for a validator to check whether it has been endorsed for selecting a next block, but also for endorsed validators to verify each other's membership. If an adversary has managed to be selected (which can be detected), a fallback mechanism needs to be employed, or simply the endorsement of a new set of validators. The complexity of this approach mainly lies in providing guarantees concerning security and reliability, both of which have been studied for decades and for which solutions exist. Using a set of endorsed validators may be the way out of blockchain's scalability problems addressed in this paper, if we can ensure that the set can remain limited (thus falling back into the advantages of traditional talking-based protocols). So far, this seems to be the case, as demonstrated for Algorand, although more work is needed.

\section{Conclusions}

When overlooking the current discussions in popular and scientific press on blockchains, we believe there is reason for serious concerns when claims are made that there is no longer need for a trusted third party while attaining scalability. Scalability in terms of high transaction capacities can be realized by concentrating validation efforts at just a few parties. Using standard distributed consensus protocols, we also arrive at highly robust solutions. The price to pay is trusting those parties. In a truly open environment with potentially thousands of participants, we are almost necessarily faced with a drop in transaction processing capacity for the simple reason that it takes time to distribute information. Reaching consensus among many also requires more: we need to have some assurance that a majority will properly follow the protocol on deciding which participant may move forward, in turn affecting the transaction processing capacity. These considerations will cause continued increase in the energy inefficiency challenges that open blockchain systems represent (million or perhaps billion-fold worse that localized trust-based systems).  Systems with computational races for reaching consensus are obviously not the way to go: apart from moral judgements on the waste of energy, their is no evidence whatsoever that they can reach high transaction processing capacities.

To what extent transaction processing capacity, scalability in participants, and openness cannot be combined remains to be seen. As of this writing, using a limited set of securely and randomly chosen endorsed validators seems to be the most promising avenue.


\end{document}